\documentclass[sigconf,natbib=true]{acmart}

\renewcommand\footnotetextcopyrightpermission[1]{}

\AtBeginDocument{%
  }

\usepackage{tabularx}
\usepackage{multirow}
\usepackage{tikz}

\begin{document}

\title{A Reproducibility Study of Graph-Based Legal Case Retrieval}

\author{Gregor Donabauer}
\affiliation{%
  \institution{University of Regensburg}
  \city{Regensburg}
  \country{Germany}
}
\email{gregor.donabauer@ur.de}

\author{Udo Kruschwitz}
\affiliation{%
  \institution{University of Regensburg}
  \city{Regensburg}
  \country{Germany}
}
\email{udo.kruschwitz@ur.de}

\renewcommand{\shortauthors}{Donabauer and Kruschwitz}

\begin{abstract}
Legal retrieval is a widely studied area in Information Retrieval (IR) and a key task in this domain is retrieving relevant cases based on a given query case, often done by applying language models as encoders to model case similarity. Recently, Tang et al. proposed CaseLink, a novel graph-based method for legal case retrieval, which models both cases and legal charges as nodes in a network, with edges representing relationships such as references and shared semantics. This approach offers a new perspective on the task by capturing higher-order relationships of cases going beyond the stand-alone level of documents. However, while this shift in approaching legal case retrieval is a promising direction in an understudied area of graph-based legal IR, challenges in reproducing novel results have recently been highlighted, with multiple studies reporting difficulties in reproducing previous findings. Thus, in this work we reproduce CaseLink, a graph-based legal case retrieval method, to support future research in this area of IR. In particular, we aim to assess its reliability and generalizability by (i) first reproducing the original study setup and (ii) applying the approach to an additional dataset. We then build upon the original implementations by (iii) evaluating the approach's performance when using a more sophisticated graph data representation and (iv) using an open large language model (LLM) in the pipeline to address limitations that are known to result from using closed models accessed via an API. Our findings aim to improve the understanding of graph-based approaches in legal IR and contribute to improving reproducibility in the field. To achieve this, we share all our implementations and experimental artifacts with the community.\footnote{\url{https://github.com/doGregor/caselink_reproducibility}}

\end{abstract}



\maketitle
\begin{tikzpicture}[remember picture,overlay]
  \node[anchor=south,yshift=30pt] at (current page.south) {
    \parbox{\textwidth}{
      \centering
      \large
      Preprint accepted at SIGIR 2025.
    }
  };
\end{tikzpicture}

\section{Introduction}

\label{sec:intro}

Legal retrieval has attracted growing attention within the Information Retrieval (IR) community over time. This becomes evident from various events and projects related to the topic\footnote{\url{https://legalai2020.github.io/}} \footnote{\url{https://trec-legal.umiacs.umd.edu/}} \cite{verberne2023ecir,coliee_2022,coliee_2023,coliee_2024}.

One specific legal IR task, legal case retrieval, 
falls within the field of legal justice and focuses on retrieving relevant cases based on a given query case \cite{shao2023understanding,feng-etal-2024-legal,shao2021investigating}. Legal experts can then analyze these cases to make informed judgments about the case in question \cite{sailer_lm_2023}.

Typically, case similarity in case retrieval is determined by using encoded representations of the individual case texts. This can be done by applying language models specifically trained on legal case data \cite{shao2020bert,sailer_lm_2023}, using large language models (LLMs) to preprocess the texts \cite{promptcase_2024}, or representing legal semantics within each case text as graphs, which are then aggregated using graph neural networks (GNNs) \cite{casegnn_2024}.

More recently, Tang et al. \cite{tang_et_al_2024} proposed an approach, called CaseLink, that goes beyond the document level by considering naturally occurring relationships between cases in a network structure. They argue that both cases and legal charges can be represented as nodes in a graph, while relationships between them (such as references, semantics or higher-order relationships) can be modeled as edges. This contextual information, which is then processed with a GNN, has the potential to uncover relationships that are not visible when cases are viewed on an individual level, and offers a novel direction in the field of legal case retrieval.

While the potential benefits of integrating contextual and structured information into legal IR have repeatedly been highlighted \cite{10.1145/3609796,fink2022graph}, there has been little research in this area so far.

Consequently, CaseLink introduces a novel perspective on legal case retrieval in an understudied area of graph-based legal IR. However, challenges related to the reproducibility of published work have been highlighted in recent years \cite{sinha2024exploring,rau2024query,koracs2024second,shehzad2024performance,bi2024reproducibility,10.1145/3539618.3591915}. To address this, we aim to contribute to the field by evaluating the reproducibility of recent research to determine whether it is reliable, referenceable, and extensible for the future.

In our work, we first reproduce the work on graph-based legal case retrieval of Tang et al. \cite{tang_et_al_2024}, presented at SIGIR 2024. We then run additional ablation studies to evaluate the generalizability of the results and assess whether extending CaseLink with other concepts of graph machine learning can positively impact the findings. 

\subsubsection*{Main Contributions.}
Our key contributions can be summarized as follows:
\begin{enumerate}
    \renewcommand{\labelenumi}{\roman{enumi}.}
    \item We conduct a reproducibility study of graph-based legal case retrieval (CaseLink) using two benchmark datasets from the COLIEE 2022 and 2023 competitions.
    \item We extend this study by including an additional dataset from COLIEE 2024 to assess whether performance remains consistent across new data.
    \item We evaluate the performance of heterogeneous graphs, which account for different node and edge types, and compare it with the homogeneous graphs used in the original work.
    \item We explore the impact of replacing GPT-3.5 (as deployed in the original study) with an open LLM in the processing pipeline.
    \item Lastly, we provide all of our implementations and artifacts to support the reproducibility of all results presented.\footnote{\url{https://github.com/doGregor/caselink_reproducibility}}
\end{enumerate}

The remainder of this paper is structured as follows: Section \ref{sec:background} provides background information on the importance of Legal IR in the broader area of Professional Search as well as on recent challenges related to reproducibility in IR. In Section \ref{sec:RQ}, we then outline our research questions before describing the methodology behind the work we reproduce in Section \ref{sec:methodology}. Sections \ref{sec:experiments} and \ref{sec:experiments_results} detail our experimental setup and findings. Finally, we discuss our results and conclusions in Section \ref{sec:discussion_conclusion}, followed by considering the paper's limitations (Section \ref{sec:limitations}) and ethical considerations (Section \ref{sec:ethical_considerations}).

\section{Background}
\label{sec:background}

\subsection{Legal Retrieval}

A major part of information searching happens in the workplace, where professionals handle large amounts of information \cite{Verberne24Professional}. Legal retrieval is a specialized domain within professional search \cite{russell2018information} and focuses on identifying and retrieving information essential for legal decision-making \cite{Barkan09Fundamentals}. It is performed by legal professionals, such as lawyers, with the primary objective of gathering evidence to answer legal questions or support specific legal positions and arguments \cite{Barkan09Fundamentals}.

One key task in this domain is legal case retrieval, which involves identifying relevant cases based on a given query case \cite{shao2023understanding,feng-etal-2024-legal,shao2021investigating}. Precedents play an important role in constructing legal arguments in common law systems \cite{shao2021investigating} and are also essential in civil law systems, where drawing analogies between relevant prior cases is necessary to ensure justice \cite{hamann2019german}.

Most research in IR has concentrated on the optimization and evaluation of ranking algorithms for web search instead of professional search, for example due to its greater commercial value \cite{Verberne24Professional}. This also applies to the legal domain and legal retrieval, which includes legal case retrieval and primarily targets legal professionals with specialized knowledge in law \cite{shao2021investigating}.

In summary, IR research has focused less on professional search compared to web search. Therefore, it is important to also concentrate on research in professional domains such as the legal domain. Explainability and transparency have been identified as key future research directions in this field \cite{Verberne24Professional}. In line with this, we aim to contribute by conducting a reproducibility study on legal case retrieval, with the goal of improving the understanding of retrieval algorithms, their reliability, and their generalizability.

\subsection{Reproducibility Issues in IR}

Recent research on reproducibility in IR highlights two challenges: (1) difficulties in reproducing the experimental results reported in the original studies and (2) assessing the generalizability of approaches beyond the datasets used in the original work.

The challenge of reproducing reported results can be observed across various IR tasks. For example, in unsupervised query generation for re-ranking computational complexity was too high to rerun experiments with a solid hardware setup \cite{rau2024query}. Similarly, in abstractive summarization with semantic graphs, performance fell below the original study’s baselines despite following the same experimental configuration \cite{koracs2024second}. This also demonstrates that even papers published at high-quality venues may lack sufficient detail for successful reproduction \cite{koracs2024second}. Furthermore, studies introducing novel methods often claim significant improvements over baselines, while recent work has shown that this is not always the case: For example, in session-based recommendation, GNN models were found to perform worse than baselines in a reproducability experiment \cite{shehzad2024performance}. In some cases, while general trends can be confirmed, reported performance metrics still show notable differences, as it was observed for a balanced topic-aware sampling method for improving PLM-based rankers \cite{10.1145/3539618.3591915}.

The second key issue, assessing the generalizability of approaches, has also been demonstrated in various contexts: For example, retrievability score calculation techniques produced different score distributions when applied across different datasets, which shows a limited robustness of these methods \cite{sinha2024exploring}. Another example is a multi-aspect dense retriever, which was evaluated on an additional dataset and performed worse than a weaker baseline \cite{bi2024reproducibility}. To even better understand generalizability, this study and other studies further analyze alternative components in the experimental setup as well as their impact on robustness and performance.

In summary, two important reproducibility challenges in IR are (1) achieving the originally reported results and (2) assessing the generalizability of approaches across datasets and experimental setups. In this work, we address these challenges by first verifying whether the original results of CaseLink can be reproduced under the same conditions and then going beyond the original experimental setup to evaluate generalizability with additional data and alternative pipeline components.

\section{Research Questions}
\label{sec:RQ}

In this paper, we formulate and address the following research questions:

\begin{itemize}
    \item \textbf{RQ1:} \textit{Are the results of CaseLink on COLIEE 2022 and 2023 reproducible?}
\end{itemize}

As an initial step, we reproduce (different team, same experimental setup\footnote{This definition is in line with the \textit{current} ACM terminology guidelines \url{https://www.acm.org/publications/policies/artifact-review-and-badging-current}}) the retrieval experiments from the original work on the two benchmark datasets from the COLIEE legal case retrieval challenges from 2022 \cite{coliee_2022} and 2023 \cite{coliee_2023}.

We face difficulties when rerunning the experiments based on the provided code, requiring communication with the authors to get the approach to work. Our findings then show quite large differences between our reproduced results and the numbers reported in the original work.

\begin{itemize}
    \item \textbf{RQ2:} \textit{Does CaseLink achieve similar performance on the COLIEE 2024 dataset?}
\end{itemize}

The results reported in the original paper show different performance outcomes between the two COLIEE datasets from 2022 and 2023. We thus extend this evaluation by including a third dataset, the one from COLIEE 2024, to assess the consistency of performance.

Our findings show that performance on COLIEE 2024 is higher than that of COLIEE 2023 but falls short of the results achieved on COLIEE 2022.

\begin{itemize}
    \item \textbf{RQ3:} \textit{How does modeling the case-charge network as a heterogeneous graph influence performance of CaseLink?}
\end{itemize}

While different document types with different relations are connected within the CaseLink graph, the original paper uses homogeneous graphs to represent the data, which do not account for these variations. We extend the setup by using heterogeneous graphs to explicitly represent the differences in node and edge types and compare the results with those from homogeneous graphs in the original work.

We find that, despite not accounting for the differences in node and edge types, homogeneous graphs result in better performance compared to heterogeneous graphs. However, these differences in performance are not significant in most cases.

\begin{itemize}
    \item \textbf{RQ4:} \textit{Does plugging in an open LLM into the CaseLink preprocessing pipeline change the overall performance?}
\end{itemize}

The complete pipeline for generating the initial CaseLink node representations is based on PromptCase \cite{promptcase_2024} and CaseGNN \cite{casegnn_2024}. As part of PromptCase, an LLM is used to generate summaries of the original cases. As previous work have highlighted reproducibility issues with models that are not publicly disclosed, we replace the GPT-3.5 model from the original work with an open LLM from the Llama family to assess whether this change results in comparable performance.

Our findings demonstrate that an open alternative shows to be a feasible solution and even results in consistently better performance, with the Llama-based CaseLink outperforming the GPT-3.5-based setup on two out of the three evaluated datasets.

\section{Methodology}
\label{sec:methodology}

Before presenting our experiments and results, we will first describe the case retrieval task in more detail and outline the methodology behind CaseLink. Additionally, we will provide a brief overview of PromptCase and CaseGNN, which are parts of the CaseLink pipeline.

\subsection{Legal Case Retrieval}

The task's starting point is a set of cases, $\mathcal{D} = \{d_{1}, d_{2}, \dots, d_{n}\}$. Given a query case $q \in \mathcal{D}$, the objective is to retrieve all cases from $\mathcal{D}$ that are relevant to $q$. This can be expressed as $\mathcal{D}^* = \{d_{i}^* | d_{i}^* \in \mathcal{D} \wedge relevant(d_{i}^*, q)\}$, where $relevant(d_{i}^*, q)$ indicates that $d_{i}^*$ is relevant to the query case $q$. In the legal domain, relevant cases are those that can serve as precedents to be referenced by the query case.

\subsection{CaseLink Graph}

The approach first constructs a homogeneous graph $G = (\mathcal{V}, \mathcal{E})$, where $\mathcal{V}$ represents the set of nodes and $\mathcal{E}$ represents the set of edges connecting these nodes. The graph contains two types of nodes: nodes representing cases $d \in \mathcal{V}$ and nodes representing charges $c \in \mathcal{V}$. Case nodes correspond to all documents from the original set of cases $\mathcal{D}$, while charge nodes are derived from a list of federal court acts\footnote{The benchmark datasets we use in our experiments are related to laws of the Federal Court of Canada, thus the charges are extracted from a list of the Federal Courts Act and Rules of Canada.} the cases are related to. This set of legal charges can be represented as $\mathcal{C} = \{c_{1}, c_{2}, \dots, c_{m}\}$. Each node is assigned a feature vector $x_{v}$, obtained using encoders that generate embeddings for the respective case or charge texts. The encoder can be any model capable of encoding case or charge texts, such as BERT \cite{devlin-etal-2019-bert} or SAILER \cite{sailer_lm_2023}. As in the original CaseLink paper we will use CaseGNN \cite{casegnn_2024} for that task.

An edge between nodes $v$ and $u$ in the graph is denoted as $e_{vu} \in \mathcal{E}$. The graph includes three types of relationships: (1) \textit{case-case} edges, (2) \textit{charge-charge} edges, and (3) \textit{case-charge} edges.

\textit{Case-case} edges connect cases that are intrinsically related and link their nodes in the graph. Pairwise BM25 scores are calculated between cases, and the $k$ most similar case pairs are added to the edge set. \textit{Charge-charge} edges model the naturally occurring relationships between different charges as multiple charges can co-occur within the same case within a legal system. Charges whose embeddings have a similarity score (calculated via dot product or cosine similarity) above a given threshold $\delta$ are connected via edges which indicates a higher likelihood of co-occurrence in similar cases. Finally, \textit{case-charge} edges link cases to the charges they contain. This can simply be determined by identifying the presence of a charge's name in the case text. All edges across these three relationships are part of the final set of edges $\mathcal{E}$.

\subsection{CaseLink Learning}

A GNN consisting of two successive graph attention layers (GAT) \cite{velivckovic2018graph} is used to aggregate information within the graph, resulting in updated node representations $\forall v \in \mathcal{V}: h_{v}^l = \text{GNN}(h_{v}^{l-1}, h_{u}^{l-1}: u \in \mathcal{N}(v))$, where $l$ represents the layer number and $h_{v}^0 = x_{v}$. To increase the expressiveness of the node embeddings, they are concatenated with their initial representations after the convolution steps via a residual connection. The method for generating these initial representations will be discussed in Section \ref{promptcase_casegnn}.

The objective is to distinguish the relevant cases from a large collection of cases based on the given query. The overall loss function is denoted as:

\begin{equation}
    l = l_{\text{InfoNCE}} + \lambda \cdot l_{\text{DegReg}}
\end{equation}

where $\lambda$ is the coefficient used to weigh the two losses. We will briefly touch on the two loss functions below, for more detail we refer to the original CaseLink paper \cite{tang_et_al_2024}.

In $l_{\text{InfoNCE}}$ a contrastive learning loss is used to distinguish between the relevant and non-relevant cases for a given query. The goal is to bring true relevant cases closer while pushing false relevant cases further away. Positive samples are taken from the ground truth, while easy negative samples are either randomly sampled or selected as hard negatives based on BM25 relevance scores.

\begin{multline}
    l_{\text{InfoNCE}} = \\
    -log\frac{e\frac{(s(h_{q},h_{d^{+}}))}{\tau}}{e\frac{(s(h_{q},h_{d^{+}}))}{\tau} + \displaystyle\sum_{i=1}^{n_{e}} e\frac{(s(h_{q},h_{d_{i}^{easy-}}))}{\tau}+\displaystyle\sum_{i=1}^{n_{h}} e\frac{(s(h_{q},h_{d_{i}^{hard-}}))}{\tau}}\label{eq:1}
\end{multline}

where $q$ is the query case, $d^{+}$ are relevant cases and $d^{easy-}$ as well as $d^{hard-}$ are easy and hard negative cases respectively. $n_{e}$ and $n_{h}$ represent the number of easy and hard negative case samples. $s$ is a similarity metric, for example cosine similarity, and $\tau$ is the temperature coefficient.

To balance the contrastive objective, which provides limited signals for candidates within the entire set of cases, an additional degree regularization loss $l_{\text{DegReg}}$ is introduced. This regularization makes sure that each candidate case is connected to only a small number of cases within the whole case set which aligns the model more closely with real-world requirements.

\begin{equation}
    l_{\text{DegReg}} = \displaystyle\sum_{i=1}^{o} \displaystyle\sum_{j=1}^{n} (\hat{A}_{ij})
\end{equation}

where $\hat{A}_{ij}$ is the pseudo adjacency matrix and $\hat{A}_{ij} = cos(h_i, h_j)$ based on the updated node features $h_i$, $h_j$ as obtained by applying the GNN to the original graph. $n$ represents all cases in the case pool in total and $o$ the number of cases in $\mathcal{D}$.

For inference the CaseLink model is applied to a graph that is constructed based on a test set of cases $\mathcal{D}_{test}$. Relevance scores $s(q,d)$ between a query case $q$ and a candidate case $d$ can then be calculated as:

\begin{equation}
    s(q, d) = \cos(h_q, h_d)
\end{equation}

where $h_q$ and $h_d$ are the representations of query case $q$ and candidate case $d$ from CaseLink. The highest scoring candidates are the ones that are retrieved.

\subsection{PromptCase and CaseGNN}
\label{promptcase_casegnn}

CaseLink uses the document representations from the baseline model CaseGNN as the initial node representations. CaseGNN relies on document representations generated by PromptCase. Thus, running these two baselines step-by-step is an important part of the whole CaseLink pipeline. For a better understanding of the pipeline and the baselines, we briefly describe both systems below. For a more detailed description of both methods we refer to the respective papers.

\subsubsection{PromptCase}

The goal of \textbf{PromptCase} \cite{promptcase_2024} is to generate more expressive case representations rather than using the raw case text directly as input to an encoder. In particular, for each case, two additional condensed versions of the original text are constructed, referred to as \textit{facts} and \textit{issues}.

For the \textit{fact} representation, the factual section of a case is provided to an LLM with the instruction to summarize it in 50 words. The \textit{issue} representation is created by extracting all sentences from the case text in which specific terms are replaced with placeholders (these are already included in the dataset), then concatenating them into a new case representation.

The \textit{facts}, \textit{issues}, and original case texts are then embedded separately using the pre-trained language model SAILER \cite{sailer_lm_2023}. Their concatenated embeddings result in a more sophisticated case representation compared to using the original case text embedding alone.

These representations can be used either for directly comparing case similarity for case retrieval or as the initial case representation for methods that build upon such embeddings, such as CaseGNN or CaseLink.


\subsubsection{CaseGNN}
\label{subsub:casegnn}

The goal of \textbf{CaseGNN} is to transform the unstructured case texts into structured text-attributed graphs and aggregate these into expressive case representations. First, relation triplets are extracted from the \textit{fact} and \textit{issue} texts that were generated by PromptCase. The extracted triplets are then used to construct separate text graphs $G_{fact}$ and $G_{issue}$ for \textit{facts} and \textit{issues} respectively. The text attributes within the graph nodes are embedded using the SAILER language model \cite{sailer_lm_2023}.

Additionally, each graph includes a virtual node that spans all other nodes within a graph (and thus can be interpreted as an overall \textit{fact} or \textit{issue} representation). This virtual node is represented by the corresponding \textit{fact} or \textit{issue} embedding generated in PromptCase. A GNN consisting of two GAT layers \cite{velivckovic2018graph} is then applied to aggregate information from the \textit{fact} and \textit{issue} graphs, producing a separate embedding for each of them. The two embeddings linked to the same case are concatenated to form a comprehensive overall representation. The learning objective is the same as presented in Equation \ref{eq:1}.

Again, these representations can be used either for directly comparing case similarity during case retrieval or as the initial case representation in CaseLink.


\section{Experimental Setup}
\label{sec:experiments}


\subsubsection*{Datasets} The original experiments were conducted using two benchmark datasets from the COLIEE 2022 \cite{coliee_2022} and COLIEE 2023 \cite{coliee_2023} legal case retrieval competitions. These datasets consist of cases collected from the Federal Court of Canada, and the training and test sets for each dataset are disconnected from each other with no overlap. Since the cases are from Canada, Tang et al. used a list of Canadian legal charges\footnote{\url{https://www.fct-cf.gc.ca/en/pages/law-and-practice/acts-and-rules/federal-court/}} in their model to create the set of charge nodes.
In addition to these two datasets, which were also used in the original paper, we include the COLIEE 2024 dataset \cite{coliee_2024}, which follows the same structure. 
We report the key statistics of the three datasets in Table \ref{table:dataset}. We note that our token counts differ from those reported in the original paper. We used the NLTK package for tokenization to calculate these statistics as the original paper did not specify the method used for their calculations.

\begin{table*}[!ht]
\begin{tabularx}{\textwidth}{|X|c|c|c|c|c|c|}
    \hline
    \multirow{2}{*}{Dataset} & \multicolumn{2}{r|}{COLIEE 2022} & \multicolumn{2}{r|}{COLIEE 2023} & \multicolumn{2}{r|}{COLIEE 2024} \\ \cline{2-7}
     & train & test & train & test & train & test \\
    \hline
    \# Queries & 898 & 300 & 959 & 319 & 1278 & 400 \\ 
    \hline
    \# Candidates & 4415 & 1563 & 4400 & 1335 & 5616 & 1734 \\
    \hline
    \# Avg. relevant cases & 4.68 & 4.21 & 4.68 & 2.69 & 4.16 & 3.91 \\
    \hline
    Avg. case length (\# tokens) & 5609.64 & 5803.21 & 5457.58 & 4855.25 & 5322.20 & 5882.63 \\
    \hline
    Largest case length (\# tokens) & 107772 & 72114 & 107772 & 52137 & 107772 & 125233 \\
    \hline
\end{tabularx}
\caption{Dataset statistics.}
\label{table:dataset}
\end{table*}


\subsubsection*{Metrics} In line with the original work we use the following standard information retrieval metrics to evaluate model performance: Precision (P), Recall (R), Micro F1 Score (Mi-F1), Macro F1 Score (Ma-F1), Mean Reciprocal Rank (MRR), Mean Average Precision (MAP), and normalized discounted cumulative gain (NDCG). We report results at $k=5$ and use the same evaluation implementations as in the original CaseLink paper.

\subsubsection*{Statistical Testing} To compare performance across different runs, we conduct statistical tests. First, we divide the test collection into five equal subsets and compute the relevant metrics at the subset level. We then apply paired t-tests with Bonferroni correction at $p<0.05$ to compare runs of multiple systems. Our evaluations are based on the NDCG@5 metric, which was used as the early stopping criterion during model training in the original study\footnote{For implementation details of the original paper, see \url{https://github.com/yanran-tang/CaseLink}}.

\label{subsec:llms}
\subsubsection*{Large Language Models} As mentioned, running CaseLink requires to first run PromptCase and CaseGNN. The LLM used in PromptCase to generate summaries of case documents is OpenAI's GPT-3.5-Turbo, which was trained for instruction-following tasks. While the exact number of parameters for this model has not been publicly disclosed, a (now-withdrawn) paper by Microsoft suggests it may have around 20 billion parameters \cite{singh2023codefusionpretraineddiffusionmodel}.
We also use Llama 3.1 \cite{dubey2024llama} from Meta AI as an open\footnote{We refer to the model as \textit{open} instead of \textit{open-source} as, while its weights are publicly available, its license does not meet open-source standards.} alternative which is available on huggingface\footnote{\url{https://huggingface.co/meta-llama/Llama-3.1-8B-Instruct}}. We again use the version of the model that was trained for instruction-following with 8 billion parameters (Llama-3.1-8B-Instruct). We will provide more motivation for selecting this specific model for our experiments in Section \ref{subsec:llama}.

\subsubsection*{Hyperparameters} For all experiments, we use the hyperparameters that are reported to result in the best performance according to the original paper. We will briefly describe them below and they are also all available on our Github.

\textbf{PromptCase} does not require setting any hyperparameter. In general, we follow the exact experimental setup from the original paper, for example for pre-processing.

For training the \textbf{CaseGNN} model, we use a learning rate of $0.000005$ with a weight decay of $0.00005$. We train for $1000$ epochs with a batch size of $32$. As mentioned in Section \ref{subsub:casegnn}, training requires using a temperature parameter and (hard) negative samples for contrastive learning. We set $\tau=0.1$ and the sample numbers to $n_{e}=1$ and $n_{h}=5$.

To construct the \textbf{CaseLink} graphs, we set $k=5$ (number of BM25-based top similar case pairs) to get the \textit{case-case} edges and $\delta = 0.9$ (embedding similarity threshold) to get the \textit{charge-charge} edges.

During model training, we set $\lambda = 0.001$ (balance between the two loss functions) and the number of negative samples $d^{-}$ to $n_{e}=1$ and $n_{h}=5$. We again set the temperature parameter $\tau=0.1$. We train with a batch size of $128$ and a learning rate of $0.00001$ (without weight decay) for $1000$ epochs. The number of GNN layers in the neural network is set to $2$.

\section{Experiments and Results}
\label{sec:experiments_results}

In this section we describe our experiments and results to the research questions proposed in Section \ref{sec:intro}. Each subsection corresponds to one of the research questions. All experiments were executed using two Nvidia RTX A6000 GPUs with 48 GB VRAM each. For comparison between runs during discussion of our findings, we primarily use the NDCG metric, as Tang et al. select the reported results based on this measure.



\subsection{Legal Case Retrieval Reproducibility}
\label{subsec:reproduced}

Our first goal is to find out whether the CaseLink results reported on the COLIEE 2022 and COLIEE 2023 datasets are reproducible. This is addressed by our first research question:

\begin{itemize}
\item \textbf{RQ1:} \textit{Are the results of CaseLink on COLIEE 2022 and 2023 reproducible?}
\end{itemize}

To answer this, we reran all the steps required to get CaseLink results on these two datasets. This also includes running both PromptCase and CaseGNN, whose results we will report as strong baselines along BM25 as IR standard baseline. Despite the authors providing their implementations\footnote{\url{https://github.com/yanran-tang/CaseLink}}, we encountered several issues while reproducing the experiments, which we will outline below.

First, the code of the three approaches is located in three different repositories. This means that we could not simply run all steps that are needed for reproducing CaseLink based on the CaseLink repository. To resolve this, we merged the methods into one shared repository with a consistent folder structure which simplifies reproducibility and makes it easier to apply the approach to new datasets in the future.

We were also facing problems when attempting to rerun the summary generation step in PromptCase, which in the original paper is based on GPT-3.5-Turbo. Since OpenAI's API model names change over time, we contacted the authors to clarify which specific model version they used during their experiments. Communication with the authors was quick and helpful, and they informed us about having used GPT-3.5-Turbo between June and December 2023. The model active during most of that time (\textit{gpt-3.5-turbo-0613}, released on 13th of June 2023) is deprecated since June 2024. We therefore used the closest model, \textit{gpt-3.5-turbo-1106}, which was released on 6th of November 2023. This shows that using commercial LLMs is not fully reproducible, as support can be discontinued by the providing companies.

Next, we encountered issues when reproducing CaseGNN (the next step of the CaseLink pipeline after PromptCase). CaseGNN requires to (1) filter case data by year and (2) identify the top 50 matching cases based on BM25. However, scripts to run these two steps were initially not provided in the respective repository. After contacting the authors, they referred us to a script from another repository that could be adapted for year filtering, which resulted in the same results as in processed files provided in their GitHub repository. We added this script to our reproducibility repository to support future experiments. However, the instructions for generating the top 50 BM25 matches were vague, and despite testing various n-gram configurations, we were unable to reproduce their exact ranking files. We instead use the most similar results we could generate. The authors noted that BM25 can have some variations, which may have influenced the discrepancies in our results.

Additionally, we found minor errors related to incorrect file and folder path names in the provided scripts, which we had to correct to ensure the pipeline ran correctly. The final step before running CaseLink -- saving the CaseGNN embeddings to be used as initial node representations for CaseLink -- was missing and we implemented a solution by ourselves.

\begin{table*}[!ht]
\begin{tabularx}{\textwidth}{|X|c|c|c|c|c|c|c|} 
    \hline
    \textbf{Method} & \textbf{P@5} & \textbf{R@5} & \textbf{Mi-F1} & \textbf{Ma-F1} & \textbf{MRR@5} & \textbf{MAP} & \textbf{NDCG@5} \\
    \hline
    BM25 & 17.7 & 21.0 & 19.2 & 21.2 & 39.8 & 38.7 & 43.1* \\
    diff. to BM25 in \cite{tang_et_al_2024} & (\textminus0.2) & (\textminus0.2) & (\textminus0.2) & (\textminus0.2) & (+16.2) & (+13.3) & (+9.5) \\
    \hline
    PromptCase & 17.5 & 20.8 & 19.0 & 21.0 & 34.2 & 33.0 & 37.9* \\
    diff. to \cite{promptcase_2024} & (+0.4) & (+0.5) & (+0.5) & (+0.5) & (\textminus0.9) & (\textminus0.9) & (\textminus0.8) \\
    \hline
    CaseGNN & \textbf{32.9} & \textbf{39.0} & \textbf{35.7} & \textbf{40.5} & \textbf{66.0} & \textbf{63.6} & \textbf{69.3} \\
    diff. to \cite{casegnn_2024} & (\textminus2.6) & (\textminus3.1) & (\textminus2.7) & (\textminus1.9) & (\textminus0.8) & (\textminus0.8) & ($\pm$0.0) \\
    \hline
    CaseLink & 30.9 & 36.7 & 33.5 & 37.5 & 61.1 & 58.4 & 64.9 \\
    diff. to \cite{tang_et_al_2024} & (\textminus6.1) & (\textminus7.2) & (\textminus6.6) & (\textminus6.7) & (\textminus6.2) & (\textminus6.6) & (\textminus5.4) \\
    \hline
\end{tabularx}
\caption{Results on the COLIEE 2022 dataset. Significant differences on the NDCG@5 metric compared to CaseLink using paired $t$-tests and Bonferroni correction at $p < 0.05$ are marked with *. Bold indicates best result for this metric.}
\label{table:results_coliee2022}
\end{table*}

\begin{table*}[!ht]
\begin{tabularx}{\textwidth}{|X|c|c|c|c|c|c|c|} 
    \hline
    \textbf{Method} & \textbf{P@5} & \textbf{R@5} & \textbf{Mi-F1} & \textbf{Ma-F1} & \textbf{MRR@5} & \textbf{MAP} & \textbf{NDCG@5} \\
    \hline
    BM25 & 16.5 & 30.6 & 21.4 & 22.1 & \textbf{41.0} & \textbf{40.3} & \textbf{44.0} \\
    diff. to BM25 in \cite{tang_et_al_2024} & ($\pm$0.0) & ($\pm$0.0) & ($\pm$0.0) & (\textminus0.1) & (+17.9) & (+19.9) & (+20.3) \\
    \hline
    PromptCase & 14.4 & 26.8 & 18.8 & 19.3 & 27.4 & 26.6 & 31.0 \\
    diff. to \cite{promptcase_2024} & (\textminus1.6) & (\textminus2.9) & (\textminus2.0) & (\textminus2.2) & (\textminus5.3) & (\textminus5.4) & (\textminus5.2) \\
    \hline
    CaseGNN & 12.8 & 23.8 & 16.6 & 16.7 & 29.2 & 28.3 & 32.5 \\
    diff. to \cite{casegnn_2024} & (\textminus4.9) & (\textminus9.0) & (\textminus6.4) & (\textminus6.9) & (\textminus9.7) & (\textminus9.4) & (\textminus10.3) \\
    \hline
    CaseLink & \textbf{17.1} & \textbf{31.8} & \textbf{22.3} & \textbf{22.7} & 34.1 & 33.3 & 38.6 \\
    diff. to \cite{tang_et_al_2024} & (\textminus3.8) & (\textminus6.6) & (\textminus4.8) & (\textminus5.5) & (\textminus11.7) & (\textminus11.0) & (\textminus11.2) \\
    \hline
\end{tabularx}
\caption{Results on the COLIEE 2023 dataset. Significant differences on the NDCG@5 metric compared to CaseLink using paired $t$-tests and Bonferroni correction at $p < 0.05$ are marked with *. Bold indicates best result for this metric.}
\label{table:results_coliee2023}
\end{table*}

\subsubsection*{Results.}

We present the reproduced results for CaseLink along with the baselines BM25, PromptCase, and CaseGNN in Table \ref{table:results_coliee2022} for the COLIEE 2022 dataset and Table \ref{table:results_coliee2023} for the COLIEE 2023 dataset. In addition, we include a comparison of the absolute differences relative to the numbers reported in the respective original papers.

One surprising observation is that while we get nearly identical results for BM25 in terms of Precision, Recall, and F1 scores, the MRR, MAP, and NDCG scores were much higher on both datasets compared to those originally reported.

For the other methods (all related to CaseLink), the results are more diverse. On the COLIEE 2022 data PromptCase and CaseGNN both result in quite similar numbers as originally reported while on the COLIEE 2023 dataset the differences between reproduced and originally reported results are quite high for both methods: For PromptCase performance is 1.6 (Precision) to 5.4 (MAP) lower, for CaseGNN the differences range from \textminus4.9 (Precision) to \textminus10.3 (NDCG).

Finally for CaseLink, which is the main interest of our reproducibility experiments, we observe a drop between 5.4 (NDCG) and 7.2 (Recall) on the COLIEE 2022 dataset and a drop of 3.8 (Precision) to 11.7 (MRR) on the COLIEE 2023 data compared to the numbers reported in the original paper.

Since PromptCase and CaseGNN are preprocessing components of CaseLink, the lower performance of CaseLink could be partially explained by the already lower performance of these two methods within the pipeline. However, on the COLIEE 2022 dataset, only CaseLink underperforms compared to the original results, while PromptCase and CaseGNN show performance that is very similar with the originally reported numbers.

\subsubsection*{Answer to RQ1.} To summarize the findings related to our first research question, we were unable to reproduce the originally reported results of CaseLink on the COLIEE 2022 and COLIEE 2023 datasets. Across both datasets, the results we got were consistently lower for all metrics compared to those reported in the original paper.

\subsection{Performance on New Data}

As observed in the results for COLIEE 2022 and 2023, CaseLink's performance variations between the two datasets were considerable (i.e. NDCG 64.9 vs. 38.6). To further assess CaseLink's generalizability, we evaluate the approach on the most recent COLIEE dataset from 2024 and want to address our second research question:

\begin{itemize}
\item \textbf{RQ2:} \textit{Does CaseLink achieve similar performance on the COLIEE 2024 dataset?}
\end{itemize}

The COLIEE 2024 dataset \cite{coliee_2024}, like those from COLIEE 2022 and 2023, is part of the same legal case retrieval competition and was released during the most recent edition. It is slightly larger than the previous datasets, as shown in the key statistics in Table \ref{table:dataset}.

As mentioned earlier, one issue of the original implementations was that they did not support datasets beyond those used in the original paper. This for example was evident in the use of hardcoded folder names within the code.

To improve the applicability of the approach on new datasets, we generalized the folder structure so that it does not include any dataset-specific subnames anymore. Additionally, we revised all file paths in the scripts to align with the updated structure and implemented automatic saving of files in a unified location. This allows the approach to be more adaptable to unseen data.

\begin{table*}[!ht]
\begin{tabularx}{\textwidth}{|X|c|c|c|c|c|c|c|}
    \hline
    \textbf{Method} & \textbf{P@5} & \textbf{R@5} & \textbf{Mi-F1} & \textbf{Ma-F1} & \textbf{MRR@5} & \textbf{MAP} & \textbf{NDCG@5} \\
    \hline
    BM25 & 18.5 & 23.7 & 20.8 & 22.0 & 42.9 & 41.4 & 45.7 \\
    \hline
    PromptCase & 18.3 & 23.4 & 20.6 & 22.6 & 30.8 & 30.0 & 34.6* \\
    \hline
    CaseGNN & 16.4 & 21.0 & 18.4 & 19.4 & 35.1 & 33.6 & 38.2* \\
    \hline
    CaseLink & \textbf{22.2} & \textbf{28.4} & \textbf{24.9} & \textbf{26.5} & \textbf{42.9} & \textbf{41.2} & \textbf{47.1} \\
    \hline
\end{tabularx}
\caption{Results on the COLIEE 2024 dataset. Significant differences on the NDCG@5 metric compared to CaseLink using paired $t$-tests and Bonferroni correction at $p < 0.05$ are marked with *. Bold indicates best result for this metric.}
\label{table:results_coliee2024}
\end{table*}

\subsubsection*{Results.}Our results on the COLIEE 2024 dataset are presented in Table \ref{table:results_coliee2024}. CaseLink achieves the highest scores across all metrics compared to the three baselines, which is different to COLIEE 2022 and 2023, where CaseGNN and BM25 respectively resulted in the highest scores (for some metrics). Looking at NDCG score differences, for COLIEE 2022, CaseLink performed 4.4 lower than the best baseline CaseGNN (however, these differences are not significant), and for COLIEE 2023, it dropped by 5.4 compared to BM25 (again, no significant difference).

In terms of absolute values, the NDCG score for COLIEE 2024 is 47.1, compared to 64.9 (+17.8) for COLIEE 2022 and 38.6 (\textminus8.5) for COLIEE 2023. This places the performance on the new dataset between the results we got for the two previous datasets.

\subsubsection*{Answer to RQ2.} To conclude, we answer the second research question with yes, CaseLink's performance on the COLIEE 2024 dataset is indeed comparable to previous results, as the observed outcomes fall between the performance levels achieved on the COLIEE 2022 and 2023 datasets.

\subsection{CaseLink with Heterogeneous Graphs}

In the original CaseLink paper Tang et al. model the data as homogeneous graphs despite representing different types of nodes (case and charge) and edges (\textit{case-case}, \textit{charge-charge}, and \textit{case-charge}). Recent research in other IR-related areas has shown that using heterogeneous graphs that explicitly account for differences in nodes and edges can improve performance, e.g. in fake news detection \cite{donabauer2023exploring}. Additionally, in a section called \textit{Graph Extensions} Tang et al. mention positive implications that could result from modelling the data as a heterogeneous graph.

We thus construct a heterogeneous case-charge graph $\mathcal{G}=(\mathcal{V}, \mathcal{E})$ that consists of a set of disjoint vertex sets $\mathcal{V} = \mathcal{V}_{D} \cup \mathcal{V}_{C}$ where $\mathcal{V}_{D} \cap \mathcal{V}_{C} = \emptyset$ as well as edges that are satisfying constraints according to the node types they link together. More specifically, we use three explicit types of edges: \textit{case-case} ($(v_{d},\tau_{DD},u_{d}) \in \mathcal{E} \to v_{d} \in \mathcal{V}_{D}, u_{d} \in \mathcal{V}_{D}$), \textit{charge-charge} ($(v_{c},\tau_{CC},u_{c}) \in \mathcal{E} \to v_{c} \in \mathcal{V}_{C}, u_{c} \in \mathcal{V}_{C}$) and \textit{case-charge} ($(v_{d},\tau_{DC},v_{c}) \in \mathcal{E} \to v_{d} \in \mathcal{V}_{D}, v_{c} \in \mathcal{V}_{C}$). The graph structure itself as well as the training procedure are the same as in the original paper. 

We apply a GNN with Heterogeneous Graph Transformer (HGT) \cite{hgt} layers to aggregate the information in the heterogeneous graph and want to answer our third research question:

\begin{itemize}
\item \textbf{RQ3:} \textit{How does modeling the case-charge network as a heterogeneous graph influence performance of CaseLink?}
\end{itemize}

\begin{table*}[!ht]
\begin{tabularx}{\textwidth}{|X|c|c|c|c|c|c|c|c|}
\hline
\textbf{Dataset} & \textbf{Graph Type} & \textbf{P@5} & \textbf{R@5} & \textbf{Mi-F1} & \textbf{Ma-F1} & \textbf{MRR@5} & \textbf{MAP} & \textbf{NDCG@5} \\
\hline
\hline
COLIEE 2022 & homogeneous & \textbf{30.9} & \textbf{36.7} & \textbf{33.5} & \textbf{37.5} & \textbf{61.1} & \textbf{58.4} & \textbf{64.9} \\
 & heterogeneous & 28.3 & 33.6 & 30.7 & 34.9 & 60.2 & 58.4 & 63.9 \\
\hline
\hline
COLIEE 2023 & homogeneous & \textbf{17.1} & \textbf{31.8} & \textbf{22.3} & \textbf{22.7} & \textbf{34.1} & \textbf{33.3} & \textbf{38.6} \\
 & heterogeneous & 14.9 & 27.6 & 19.3 & 19.5 & 31.3 & 30.9 & 35.4 \\
\hline
\hline
COLIEE 2024 & homogeneous & \textbf{22.2} & \textbf{28.4} & \textbf{24.9} & \textbf{26.5} & \textbf{42.9} & \textbf{41.2} & \textbf{47.1}* \\
 & heterogeneous & 19.8 & 25.4 & 22.2 & 23.2 & 39.4 & 37.4 & 42.9* \\
\hline
\end{tabularx}
\caption{CaseLink results with heterogeneous graphs. Significant differences on the NDCG@5 metric between the homogeneous and heterogeneous setup for each year using paired $t$-tests at $p < 0.05$ are marked with *. Bold indicates best result for this metric within the same dataset.}
\label{table:results_heterogeneous}
\end{table*}

\subsubsection*{Results.} We present the results with heterogeneous graphs in Table \ref{table:results_heterogeneous}. As observable, all of the metrics result in the highest scores with homogeneous graphs. However, for COLIEE 2022 and 2023, the differences between homogeneous and heterogeneous graphs are not statistically significant, while for COLIEE 2024, the homogeneous setting outperforms the heterogeneous one. The absolute differences for NDCG on heterogeneous graphs amount to \textminus1.0 (COLIEE 2022), \textminus3.2 (COLIEE 2023), and \textminus4.2 (COLIEE 2024).

\subsubsection*{Answer to RQ3.} In summary, our findings suggest that heterogeneous graphs do not result in advantages for CaseLink, despite prior research indicating so. However, the overall differences between the two setups are not significant on two out of three datasets.

\subsection{Llama as LLM for Preprocessing}
\label{subsec:llama}

As mentioned in Section \ref{subsec:reproduced}, using a commercial LLM as part of PromptCase resulted in challenges for reproducibility. Previous studies have also highlighted that many models that are not publicly disclosed and offer interaction only via API hinder reproducibility \cite{balloccu2024leak,staudinger2024reproducibility}. We therefore want to evaluate how using an open alternative performs in context of CaseLink which is why we adopt Llama-3.1-8B-Instruct for the same task as outlined earlier (Section \ref{subsec:llms}).

We select Llama-3.1-8B-Instruct for two main reasons: (1) This model, along with the entire Llama-3.1 family, was released in July 2023\footnote{Compare model release date on \url{https://huggingface.co/meta-llama/Llama-3.1-8B-Instruct} or \url{https://ai.meta.com/blog/meta-llama-3-1/}}, which is in line with the release dates of the GPT-3.5-Turbo versions that were used in the original experiments; (2) As mentioned previously, OpenAI's GPT-3.5-Turbo is estimated to have around 20 billion parameters. The Llama-3.1 family includes models of varying sizes -- 8 billion, 70 billion, and 405 billion parameters. Among these, the 8 billion parameter model we adapt in our experiments is the closest in scale to the used GPT-3.5 models. Llama-3.1-8B-Instruct thus serves as a comparable open alternative to the original GPT-3.5-Turbo model.

The final research question we want to answer is:

\begin{itemize}
\item \textbf{RQ4:} \textit{Does plugging in an open LLM into the CaseLink preprocessing pipeline change the overall performance?}
\end{itemize}

\begin{table*}[!ht]
\begin{tabularx}{\textwidth}{|X|c|c|c|c|c|c|c|} 
    \hline
    \textbf{Dataset} & \textbf{P@5} & \textbf{R@5} & \textbf{Mi-F1} & \textbf{Ma-F1} & \textbf{MRR@5} & \textbf{MAP} & \textbf{NDCG@5} \\
    \hline
    \textbf{COLIEE 2022} & \multicolumn{7}{c|}{}\\
    \hline
    PromptCase with Llama & \textbf{18.7} & \textbf{22.3} & \textbf{20.3} & \textbf{22.0} & \textbf{37.0} & \textbf{35.2} & \textbf{40.4} \\
    with ChatGPT 3.5 & 17.5 & 20.8 & 19.0 & 21.0 & 34.2 & 33.0 & 37.9 \\
    \hline
    CaseGNN with Llama & \textbf{36.0} & \textbf{42.8} & \textbf{39.1} & \textbf{44.0} & \textbf{68.5} & \textbf{66.1} & \textbf{71.1} \\
    with ChatGPT 3.5 & 32.9 & 39.0 & 35.7 & 40.5 & 66.0 & 63.6 & 69.3 \\
    \hline
    CaseLink with Llama & \textbf{31.5} & \textbf{37.5} & \textbf{34.2} & \textbf{38.3} & \textbf{63.0} & \textbf{61.4} & \textbf{67.3} \\
    with ChatGPT 3.5 & 30.9 & 36.7 & 33.5 & 37.5 & 61.1 & 58.4 & 64.9 \\
    \hline
    \textbf{COLIEE 2023} & \multicolumn{7}{c|}{}\\
    \hline
    PromptCase with Llama & \textbf{15.4} & \textbf{28.5} & \textbf{20.0} & \textbf{20.6} & \textbf{29.3} & \textbf{28.7} & \textbf{32.7} \\
    with ChatGPT 3.5 & 14.4 & 26.8 & 18.8 & 19.3 & 27.4 & 26.6 & 31.0 \\
    \hline
    CaseGNN with Llama & \textbf{14.2} & \textbf{26.3} & \textbf{18.4} & \textbf{18.7} & \textbf{32.1} & \textbf{30.2} & \textbf{35.0} \\
    with ChatGPT 3.5 & 12.8 & 23.8 & 16.6 & 16.7 & 29.2 & 28.3 & 32.5 \\
    \hline
    CaseLink with Llama & \textbf{18.3} & \textbf{33.9} & \textbf{23.7} & \textbf{24.4} & \textbf{41.4} & \textbf{39.2} & \textbf{44.0}* \\
    with ChatGPT 3.5 & 17.1 & 31.8 & 22.3 & 22.7 & 34.1 & 33.3 & 38.6* \\
    \hline
    \textbf{COLIEE 2024} & \multicolumn{7}{c|}{}\\
    \hline
    PromptCase with Llama & \textbf{20.2} & \textbf{25.8} & \textbf{22.6} & \textbf{24.7} & \textbf{35.0} & \textbf{33.8} & \textbf{39.0} \\
    with ChatGPT 3.5 & 18.3 & 23.4 & 20.6 & 22.6 & 30.8 & 30.0 & 34.6 \\
    \hline
    CaseGNN with Llama & \textbf{18.4} & \textbf{23.6} & \textbf{20.7} & \textbf{21.8} & \textbf{37.7} & \textbf{36.3} & \textbf{41.8}* \\
    with ChatGPT 3.5 & 16.4 & 21.0 & 18.4 & 19.4 & 35.1 & 33.6 & 38.2* \\
    \hline
    CaseLink with Llama & \textbf{23.3} & \textbf{29.8} & \textbf{26.2} & \textbf{28.5} & \textbf{44.9} & \textbf{43.5} & \textbf{49.7}* \\
    with ChatGPT 3.5 & 22.2 & 28.4 & 24.9 & 26.5 & 42.9 & 41.2 & 47.1* \\
    \hline
\end{tabularx}
\caption{Results on the COLIEE 2022, 2023 and 2024 dataset with Llama plugged into the full pipeline. Significant differences on the NDCG@5 metric between the ChatGPT 3.5 and Llama setup within the same year using paired $t$-tests at $p < 0.05$ are marked with *. Bold indicates best result for this metric within the same dataset and approach.}
\label{table:results_llama}
\end{table*}

\subsubsection*{Results.} The results using Llama-3.1 in place of GPT-3.5-Turbo across all three COLIEE datasets are presented in Table \ref{table:results_llama}. As we can see, the scores are consistently higher across all metrics and datasets when using the open alternative compared to the commercial LLM.

Although most differences are not statistically significant, we find that CaseLink, when combined with Llama, outperforms the GPT-3.5-based setup on the COLIEE 2023 and 2024 datasets. Similarly, CaseGNN achieves significantly better performance on the COLIEE 2024 dataset with Llama than with GPT-3.5.

\subsubsection*{Answer to RQ4.} In response to our final research question, we find that using Llama-3.1-Instruct-8B as an open alternative to GPT-3.5-Turbo does indeed affect the overall performance of CaseLink. The impact is positive, as the Llama-based setup consistently results in higher scores across all evaluation metrics, for CaseLink even in significant improvements on two out of three datasets.

\section{Discussion and Conclusion}
\label{sec:discussion_conclusion}

In this reproducibility study, we evaluated a recent legal case retrieval method called CaseLink, which models contextual information between cases and charges in graphs. We encountered a number of issues when reproducing the results reported in the original experiments, which is in line with findings from the previous reproducibility tracks, where many papers reported difficulties in achieving the same results as the original studies they reproduced \cite{sinha2024exploring,rau2024query,koracs2024second,shehzad2024performance,bi2024reproducibility,10.1145/3539618.3591915}.

Additionally, we conducted ablation studies on the new COLIEE 2024 dataset to assess the generalizability of the approach. The results are between those reported on datasets from previous COLIEE competitions, suggesting a consistent but varying performance across different data. We also explored alternative graph data setups, comparing heterogeneous networks with the homogeneous graphs used in the original work. However, these more complex data representations did not result in any improvements. Furthermore, we tested an open LLM alternative in the CaseLink pipeline as part of PromptCase, addressing the challenges we faced with the commercial OpenAI GPT model used in the original paper. Our findings suggest that Llama is a viable alternative, supporting the reproducibility of LLM-based methods in the field and even resulting in consistently better performance.

We share our repository that we optimized for reproducibility of the approach with the community and hope that the insights from our experiments will support future research in IR.

\section{Limitations}
\label{sec:limitations}

While we expanded on the experimental setup to assess the reliability and generalizability of the original work, several limitations remain and we will briefly discuss them in this section.

First, our experiments focus exclusively on legal case retrieval, which is a specialized subfield within the broader domains of legal retrieval and professional search. As a result, the insights we gained from our study may not generalize to other areas of legal information retrieval. 

Second, the three benchmark datasets used in our experiments consist solely of cases from the Canadian Federal Court. This limits the applicability of the findings to other legal systems, as it remains unclear how well the approach would transfer to datasets from different jurisdictions, for example from Germany or China.

Third, regarding the graph representation we use, we have taken an initial step towards more sophisticated modeling. However, several directions for further improvements remain. For example, the graph edges could be directed based on BM25 scores, or these scores could be included as edge weights. Such work can be based on the heterogeneous graph representations introduced in our study as they make it easier to model this information.

Finally, our work does not account for the dynamic nature of legal data. The datasets and data modeling used in this study do not capture how legal cases and references evolve over time, which is an important aspect of real-world case law. Future research should explore approaches that consider temporal dynamics.

\section{Ethical Considerations}
\label{sec:ethical_considerations}

We do not identify any immediate ethical concerns arising from our work. The datasets used in our study are made available for research purposes under the condition of signing a memorandum, which we did before starting the experiments. Similarly, our use of the Llama LLM is in line with the model’s license. To make our work transparent, we publicly make available our implementations and experimental artifacts to the community, to support the reproducibility of our results.

From a broader perspective, we acknowledge that while legal retrieval tools can help in reducing the workload of legal professionals, they may also influence decision-making and critical legal reasoning. Over-reliance on AI-driven retrieval could shift responsibility away from human legal experts, which raises concerns about accountability in case selection and legal interpretation.


\bibliographystyle{ACM-Reference-Format}
\bibliography{references}


\end{document}